\documentclass[twoside,12pt]{article}
\usepackage{epsfig}

\newcommand{\be}{\begin{equation}}
\newcommand{\ee}{\end{equation}}
\newcommand{\bea}{\begin{eqnarray}}
\newcommand{\eea}{\end{eqnarray}}

\begin{document}

\title{Event rates for WIMP detection }

\author{J.D. Vergados$^{*}$, Ch.C. Moustakidis$^{\dagger}$ and V. Oikonomou$^{\dagger}$ \\
$^{*}$University of Ioannina, Gr 45110, Greece \\ $^{\dagger}$
Department of
Theoretical Physics, Aristotle University of Thessaloniki, \\
54124 Thessaloniki, Greece }

\maketitle

\begin{abstract}
The event rates for the direct detection of  dark matter for
various types of WIMPs are presented. In addition to the
neutralino of SUSY models, we considered other candidates (exotic
scalars as well as particles in Kaluza-Klein and technicolour
theories) with masses in the TeV region. Then one finds reasonable
branching ratios to excited states. Thus the detection of the WIMP
can be made not only by recoil measurements, but by  measuring the
de-excitation $\gamma$-rays as well.

\end{abstract}
\

\section{Introduction}
It is now established that the universe is dominated by dark
matter and dark energy. The nature of dark matter can only be
established if its constituents are detected in the laboratory.
There are currently many models which supply dark matter
candidates, called generically WIMPs (weakly interacting massive
particles). The candidate, most favored in R-parity conserving
supersymmetry, is the LSP (lightest supersymmetric particle).
Other candidates arise in other extensions of the standard model
and, in particular, in models with compact extra dimensions. In
such models  a tower of massive particles appear as Kaluza-Klein
excitations.
In this scheme the ordinary particles are associated with the zero
modes and are assigned K-K parity $+1$.
In models with Universal Extra Dimensions one can have
cosmologically stable particles in the excited modes because of a
discrete symmetry yielding  K-K parity $-1$ (see Servant
\cite{SERVANT} for a recent review).

The kinematics involved is similar to that of the neutralino,
leading to cross sections which are proportional $\mu^2_r$,
$\mu_r$ being the WIMP-nucleus reduced mass. Furthermore the
nuclear physics input is independent of the WIMP mass, since for
heavy WIMP $mu_r\simeq Am_p$. There appear two differences
compared to the neutralino, though,  both related to its larger
mass.

 First the density
  (number of particles per unit volume) of a WIMP
 falls inversely proportional to its mass. Thus,
  if the WIMP's considered are much heavier than the nuclear
 targets, the corresponding event rate takes the form:
  \begin{equation}
 R(m_{WIMP})=R(A)\frac{A \mbox{ GeV
}}{m_{WIMP}}
 \label{eq:rate}
 \end{equation}
 where $R(A)$ are the rates extracted from experiment up to WIMP
 masses of the order of the mass of the target.

Second, the average WIMP energy is now  higher. In fact, one finds
that $\prec T_{WIMP}\succ =\frac{3}{4}M_{WIMP} \upsilon^2_0\simeq
40 \left (\frac{m_{WIMP}}{100 \mbox{ GeV}}\right )$keV
($\upsilon_0\simeq 2.2\times 10^5$km/s).
  Thus for a K-K WIMP with mass $1$ TeV, the average
WIMP energy is $0.4$ MeV. Hence, due to the high velocity tail of
the velocity distribution,  one expects {\bf an energy transfer to
the nucleus  in the MeV region. Thus many nuclear
 targets can now be excited by the WIMP-nucleus interaction and the de-excitation photons
 can be detected.}
 \section{ Kaluza-Klein WIMP's}
 \subsection{The Kaluza-Klein Boson as a dark matter candidate}
 \label{KK}
 We will assume that the lightest exotic particle, which can serve as a
 dark matter candidate, is a gauge boson $B^{1}$
 having the same quantum numbers and couplings with the standard model gauge
  boson $B$, except that it has K-K parity
 $-1$. Thus its couplings involve another negative K-K parity particle.
 In this work we will assume that such  particles
 are the K-K quarks, partners of the ordinary quarks,
 but much heavier \cite{ST02}.
\subsubsection{Intermediate K-K quarks}
\begin{figure}[!ht]
 \begin{center}
\includegraphics[scale=0.6]{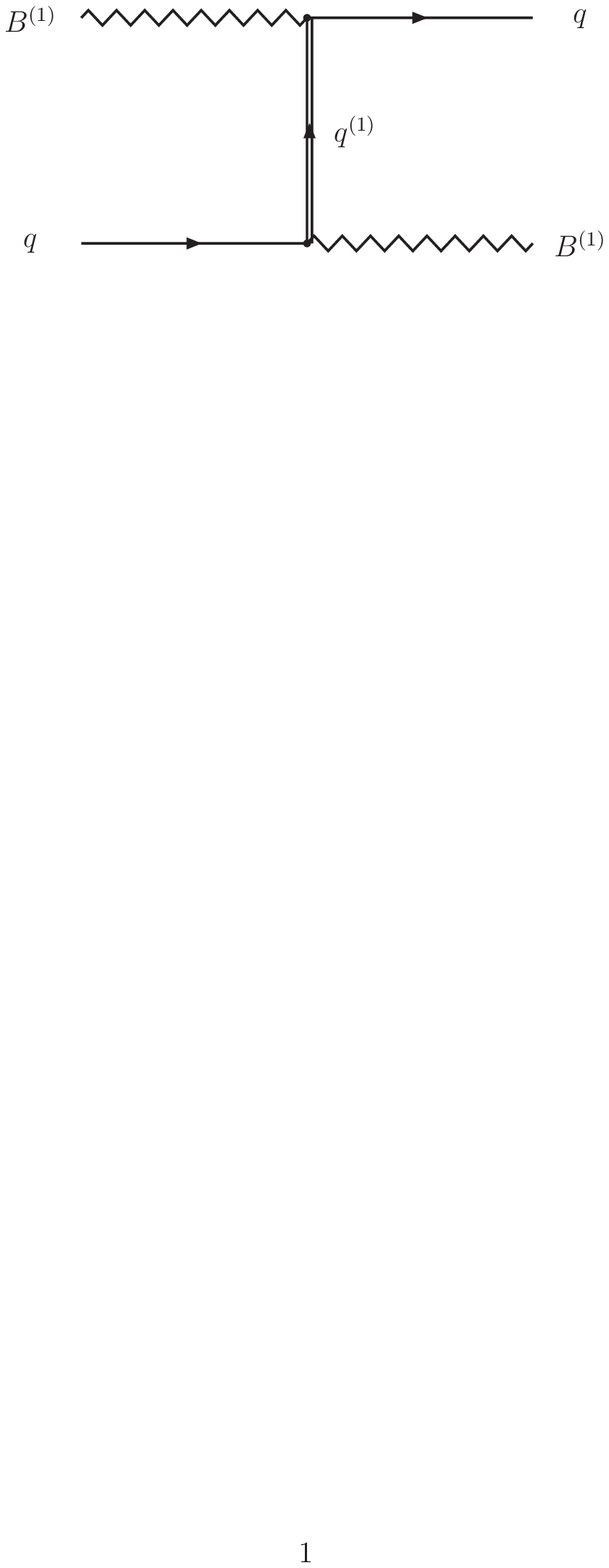}
\ \ \ \ \ \ \
\includegraphics[scale=0.6]{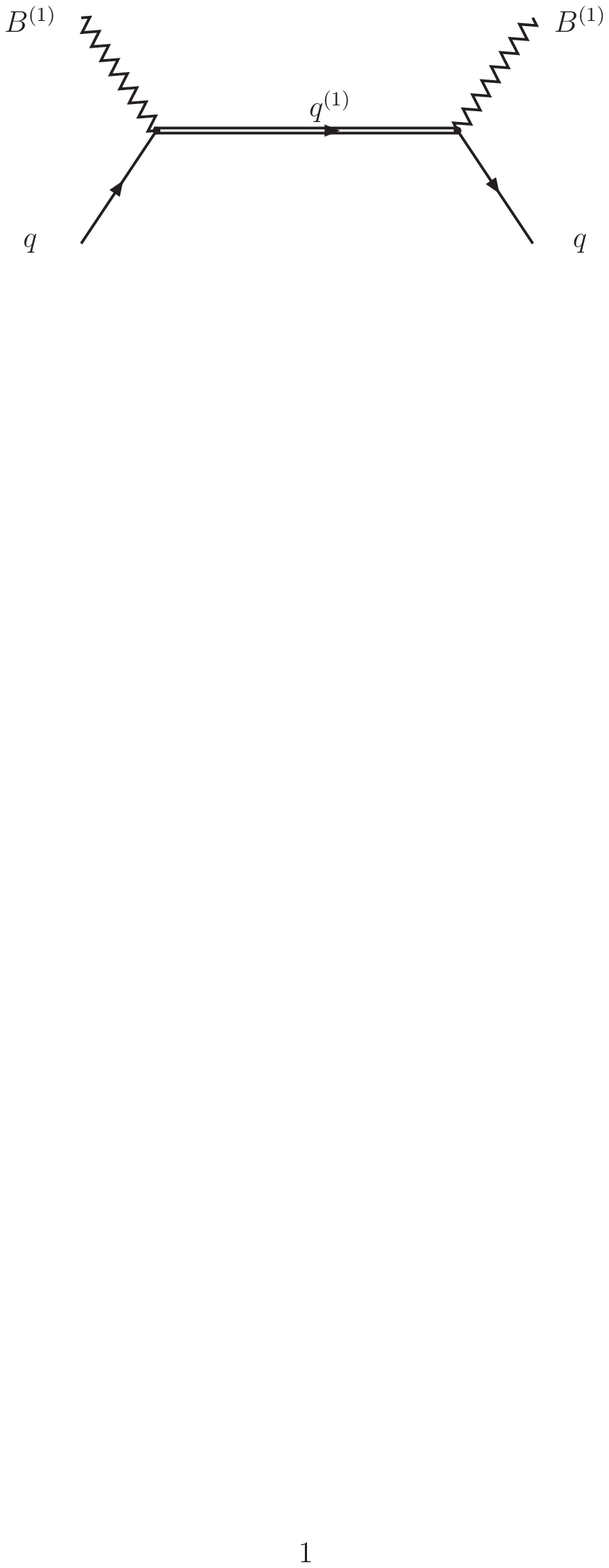}
 \caption{K-K quarks mediating the interaction of K-K gauge boson $B^{1}$
 with quarks at tree
level.}
 \label{fig:kkq}
 \end{center}
  \end{figure}

 In this case the relevant
Feynman diagrams are
 shown in fig. \ref{fig:kkq}.
The amplitude at the nucleon level can be written as:
 \begin{equation} {\cal M}_{coh}= \Lambda(\mbox{\boldmath
$\epsilon^{*'}$}.\mbox{\boldmath $\epsilon$})N\left [ \left (
\frac{11}{18}+\frac{2}{3}\tau_3 \right ) \frac{1}{3} \frac{m_p
m_W}{(m_{B^{(1)}})^2} f_1(\Delta )+\frac{1+\tau_3}{3} \frac{m_W}{
m_{B^{(1)}}} f_2(\Delta ) \right ] N \end{equation}
$$\Lambda=i 4 \sqrt{2} G_F m_W \tan^2{\theta_W
},f_1(\Delta )=\frac{1+\Delta +\Delta ^2 /2}{\Delta ^2(1+\Delta
/2)^2}, f_2 (\Delta )=\frac{1+\Delta }{\Delta (1+\Delta /2)}~,
~\Delta =\frac{m_{q^{(1)}}}{m_{B^{(1)}}}-1$$ We see that the
amplitude is very sensitive to the parameter $\Delta $ ("resonance
effect").

In going from the quark to the nucleon level the best procedure is
to replace the quark energy by the constituent quark mass $\simeq
1/3m_p$, as opposed to adopting \cite{ST02} a procedure related to
the current mass encountered in the neutralino case \cite{JDV06}.


In the case of the spin contribution we find at the nucleon level
that\begin{equation} {\cal M}_{spin}= -i 4 \sqrt{2} G_F m_W
\tan^2{\theta_W }\frac{1}{3} \frac{m_p m_W}{(m_{B^{(1)}})^2}
f_1(\Delta ) i(\mbox{\boldmath $\epsilon^{*'}$}\times
\mbox{\boldmath $\epsilon$}).\left [ N\mbox{\boldmath $\sigma$}
(g_0+g_1 \tau_3) N \right ] \end{equation}
$$g_0=\frac{17}{18}\Delta u+\frac{5}{18} \Delta d+\frac{5}{18} \Delta s~,
~g_1=\frac{17}{18}\Delta u-\frac{5}{18} \Delta d$$
for the isoscalar and isovector quantities \cite{JDV06}. The
quantities $\Delta_q$ are given by \cite{ST02}, \cite{JDV06}
$$\Delta u=0.78\pm 0.02~,~\Delta d=-0.48\pm 0.02~,~\Delta s=-0.15\pm 0.02$$
We thus find $g_0=0.26~,~g_1=0.41\Rightarrow
a_p=0.67~,~a_n=-0.15$.
The picture is different for the neutralino case
($a_p=1.41~,~a_n=-1.11$)
\subsubsection{Intermediate Higgs Scalars} 
\begin{figure}[!ht]
\begin{center}
\includegraphics[scale=0.8]{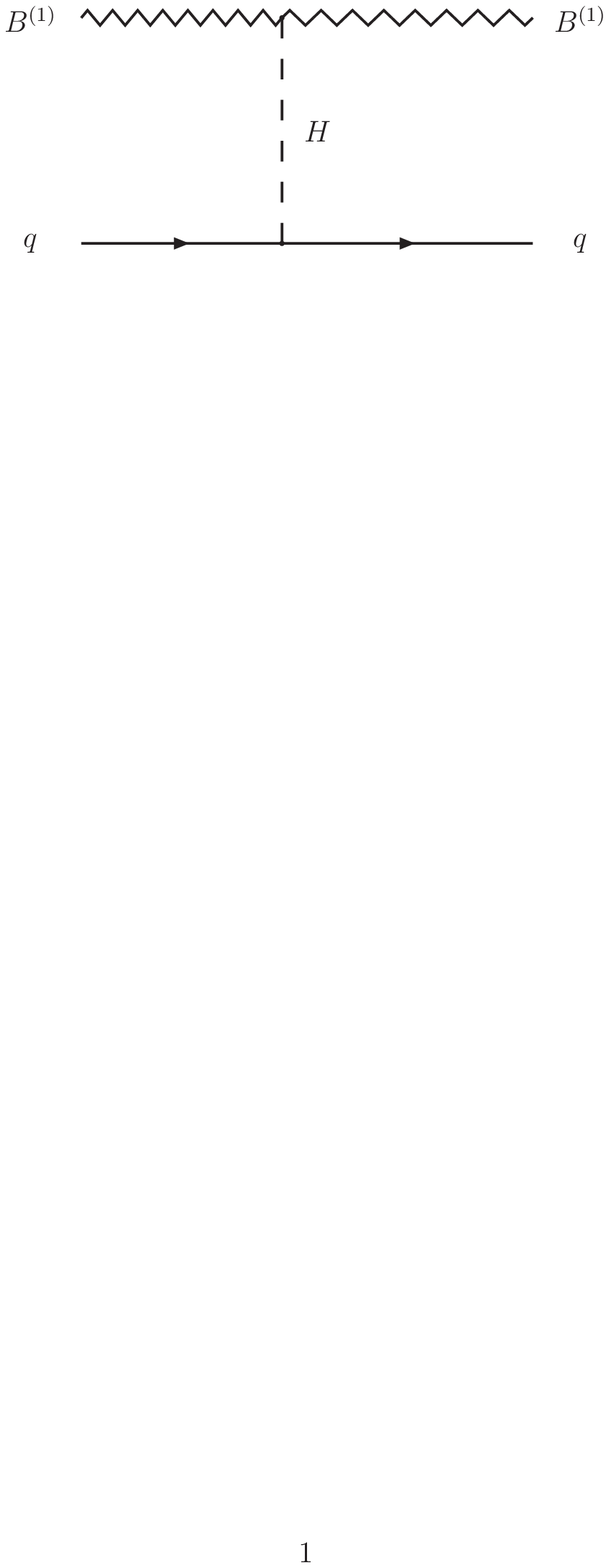}
\ \ \ \ \ \
\includegraphics[scale=0.8]{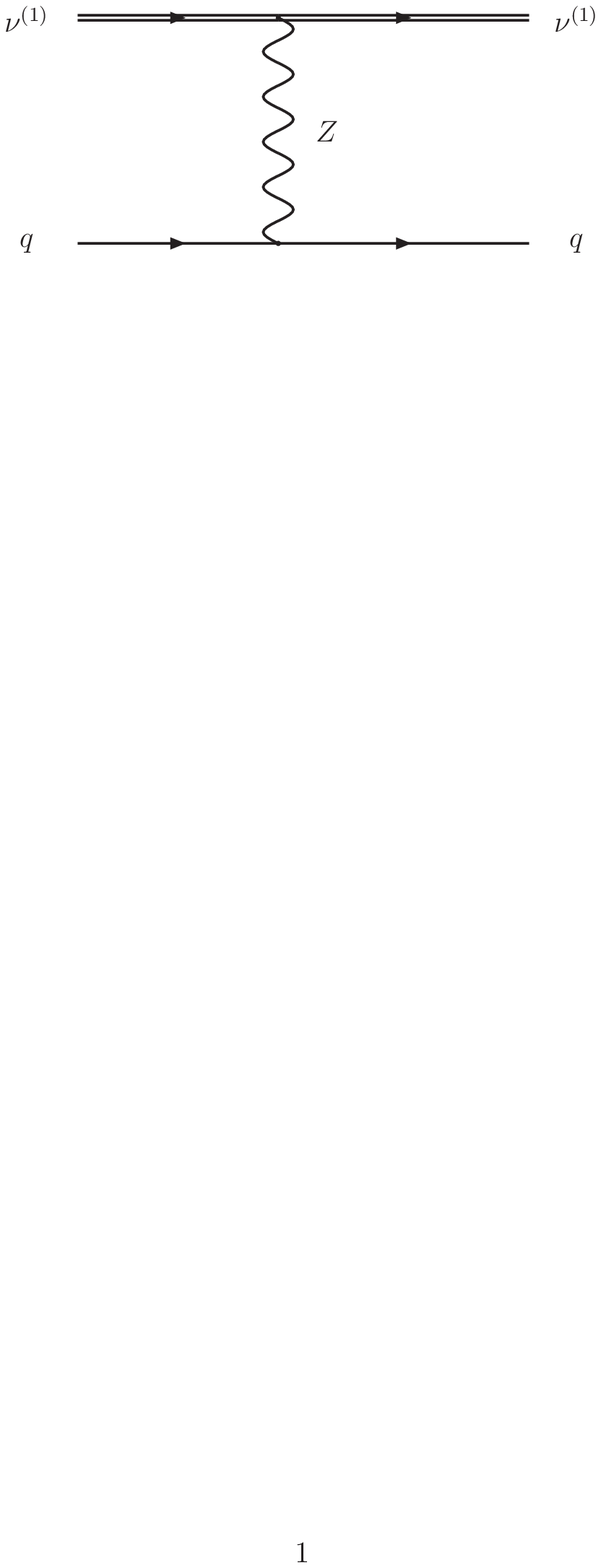}
 \caption{The Higgs H mediating interaction of K-K gauge boson $B^{1}$ with quarks at tree
level (on the left). The Z-boson mediating the interaction of K-K
neutrino $\nu^{(1)}$ with quarks at tree level (on the right).}
 \label{fig:kkhz}
\end{center}
  \end{figure}

The corresponding
Feynman diagram is shown in Fig. \ref{fig:kkhz}
The relevant amplitude is given by:
  \begin{equation}
 {\cal M}_N(h)= -i~4 \sqrt{2}G_F m^2_W \tan^2{\theta_W}~\left [\frac{1}{4}\frac{m_p}{m^2_h}
 \left (-\mbox{\boldmath
$\epsilon^{*'}$}.\mbox{\boldmath $\epsilon$} \right ) \prec
N|N\succ  \sum_q f_q\right ]
  \end{equation}
   In going from the quark to the nucleon level we follow a procedure analogous to that
 of the of the neutralino,
  i.e.
  $\prec  N |m_q q \bar{q}|N  \succ  \Rightarrow f_q m_p$
  \subsection{K-K neutrinos as dark matter candidates}
  The other possibility is the dark matter candidate to be a heavy K-K neutrino.
   We will distinguish the following cases:
\subsubsection{Process mediated by Z-exchange}
 The Feynman diagram associated
 with this process is shown in Fig. \ref{fig:kkhz}.
 The amplitude associated with the diagram of Fig. \ref{fig:kkhz} becomes:
 \begin{equation}
 {\cal M}_{\nu^{(1)}}=\frac{1}{4}\frac{g^2}{4 \cos^2{\theta_W}}\frac{1}{-m^2_Z}J^{\lambda}(\nu^{(1)}) J_{\lambda}(NNZ)=
-\frac{1}{2 \sqrt{2}}G_FJ^{\lambda}(\nu^{(1)}) J_{\lambda}(NNZ)
 \end{equation}
 with $J_{\lambda}(NNZ)$ the standard nucleon neutral current and
$$J_{\lambda}(\nu^{(1)})= \bar{\nu}^{(1)}\gamma _{\lambda
}\gamma_5\nu^{(1)}~,~J_{\lambda}(\nu^{(1)})= \bar{\nu}^{(1)}\gamma
_{\lambda }(1-\gamma_5)\nu^{(1)}$$
 for Majorana and Dirac neutrinos respectively.
 \subsubsection{Process mediated by right handed currents  via $Z'$-boson exchange}
The process is similar to that exhibited by fig. \ref{fig:kkhz},
except that instead of Z we encounter $Z'$, which is much heavier.
We will assume that the couplings of the $Z'$ are similar to those
of $Z$. Then the above results apply except that now the
amplitudes are retarded by the multiplicative factor
$\kappa=m^2_{Z}/m^2_{Z'}$
 \subsubsection{Process mediated by Higgs exchange}
 In this case in fig. \ref{fig:kkhz} the Z is replaced by the Higgs particle.
 Proceeding as above we find that the amplitude at the nucleon level is:
  \begin{equation}
 {\cal M}_{\nu^{(1)}}(h)=
-2 \sqrt{2}G_F \frac{m_p m_{\nu^{(1)}}}{m_h^2}
\bar{\nu}^{(1)}~\nu^{(1)} \prec N|N \succ \sum_q f_q
 \end{equation}
  In the evaluation of the parameters
$f_q$ one encounters both theoretical and experimental errors.
In the present calculation we will adopt an optimistic approach
and employ \cite{JDV06}:
$$f_d=0.041,~f_u=0.028,~f_s=0.400,~f_c=0.051,~f_b=0.055,~f_t=0.095$$
  \section{Nucleon cross sections}
  In evaluating the nucleon cross section one proceeds as in the case of the neutralino.
\subsubsection{The K-K boson case}
\begin{figure}[!ht]
\begin{center}
\includegraphics[scale=0.6]{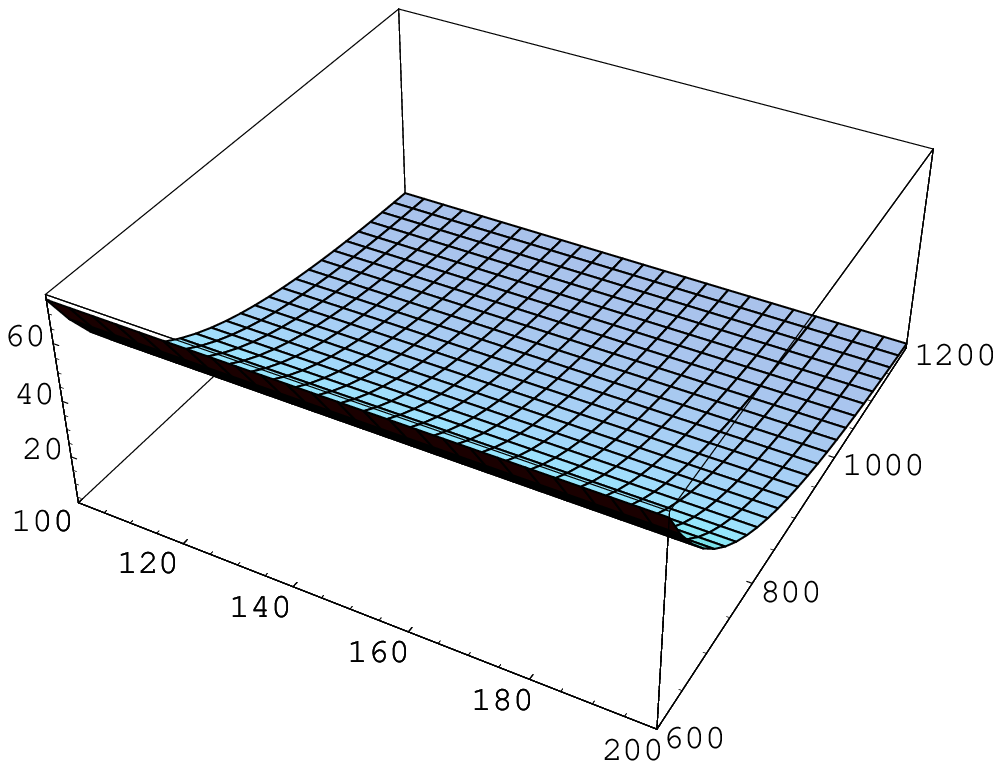}
\ \ \ \ \
\includegraphics[scale=0.6]{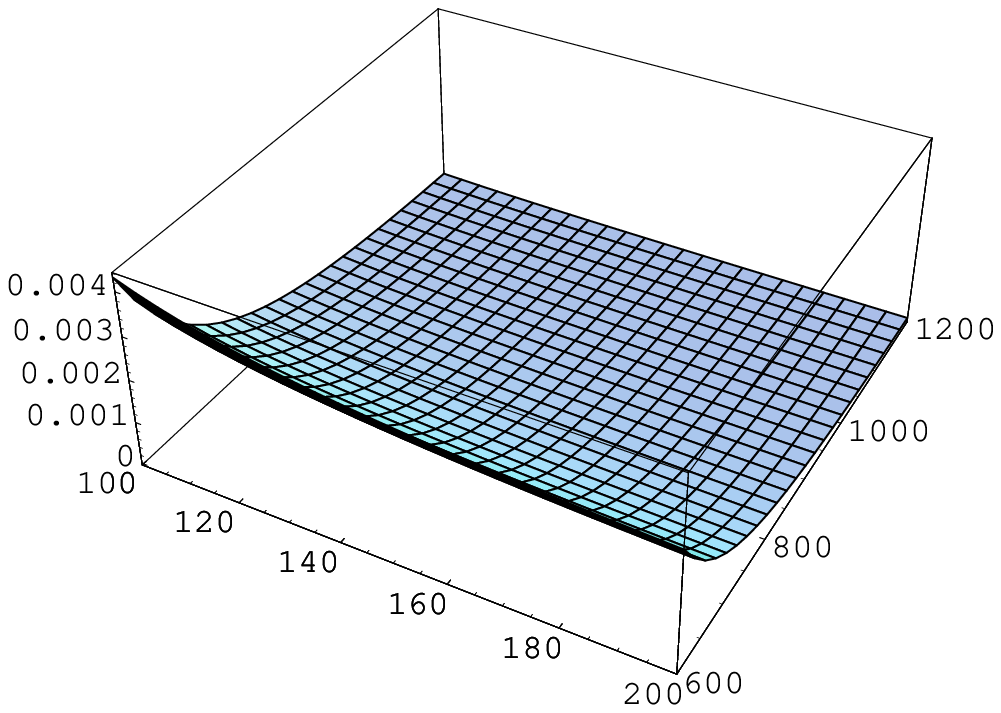}
 \caption{The coherent proton cross section on the left and that for the neutron on the right in units of
$10^{-6}$pb, as a function of the gauge boson mass in the range of
$600-1200$ GeV and the Higgs mass in the range of $100-200$ GeV
for  $\Delta=0.15$. For smaller values of $\Delta$ the proton
cross section is huge.}
 \label{fig3d:coh}
\end{center}
  \end{figure}

\begin{figure}[!ht]
\begin{center}
 \includegraphics[scale=0.6]{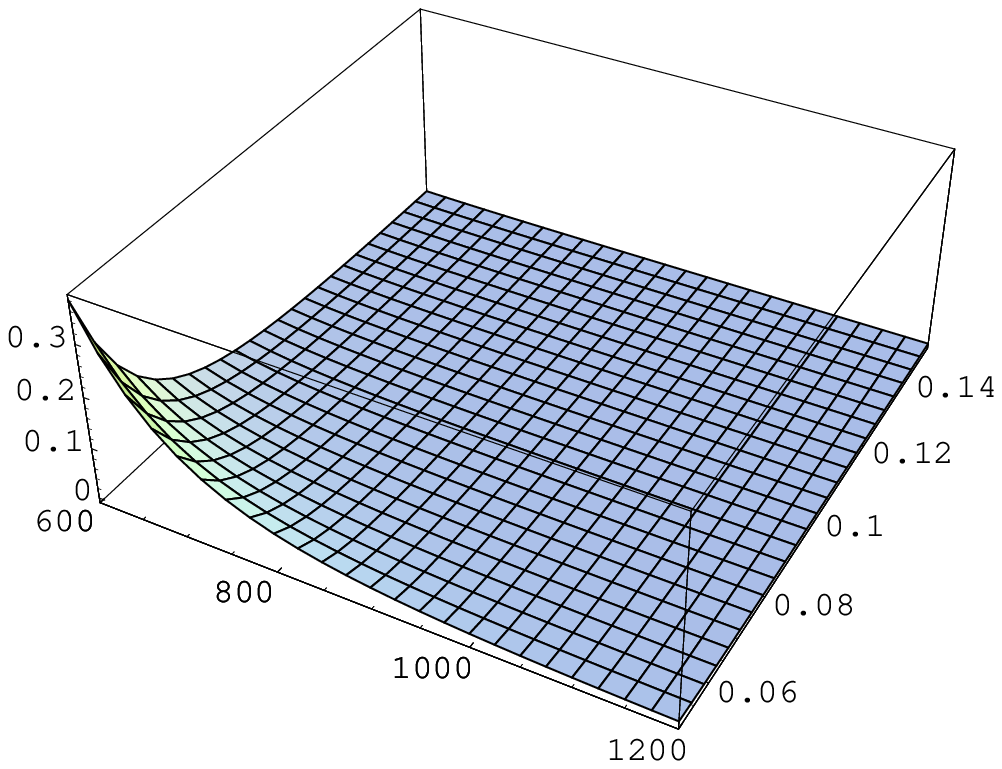}
 \ \ \ \ \ \
\includegraphics[scale=0.6]{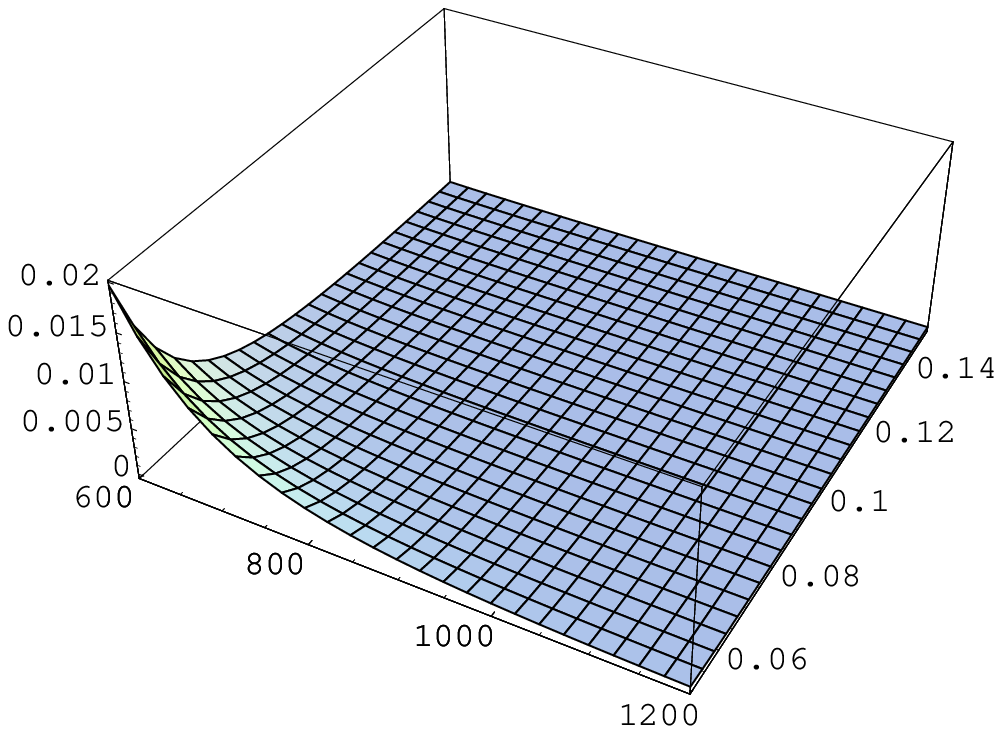}
 \caption{The spin proton cross section on the left and that for the
 neutron on the right in units of
$10^{-6}$pb, as a function of the gauge boson mass in the range of
$600-1200$ GeV and $\Delta$ in the range $0.05-0.15$.}
 \label{fig3d:spin}
\end{center}
  \end{figure}

The kinematics are similar to those of the neutralino case. One
finds
 \begin{equation}
 \sigma_N(i)=\frac{1}{4 \pi}\frac{m_p^2}{(m_{B^{(1)}})^2} \frac{1}{2} \frac{1}{3}
 \sum_{pol,m_s}|{\cal M}_{i}|^2, ~~i=coh,spin
 \end{equation}
%
 for the spin independent and spin dependent parts respectively.
The  obtained results for the coherent process are shown in fig.
\ref{fig3d:coh}
  The Higgs contribution is negligible in this case.
 In the case of the spin cross section the obtained results are shown in
 fig. \ref{fig3d:spin}.
  \subsubsection{The K-K neutrino case}
  In this case the expression for the cross section is quite simple. We will consider each
  case separately.
 \begin{enumerate}
   \item Intermediate $Z$ boson. In this case
   \begin{equation}
  \sigma_N(spin)=\frac{1}{\pi}\frac{G_F^2}{8} m^2_p 3 g_A^2=8.0\times
  10^{-3}pb \mbox{ (Majorana neutrino)}
  \end{equation}
    \begin{equation}
  \sigma_n(spin)= \sigma_p(spin)\simeq \frac{1}{\pi}\frac{G_F^2}{8} m^2_p ~3~ 2~ g_A^2
  =1.6\times 10^{-2}pb \mbox{ (Dirac neutrino)}
  \end{equation}
    \begin{equation}
  \sigma_n(coh)=\frac{1}{\pi}\frac{G_F^2}{8} m^2_p ~ 2~=3.5\times 10^{-3}pb
  \mbox{ (only Dirac neutrino)}
  \end{equation}
  It is quite straightforward to compute the nuclear cross sections:
  \begin{equation}
  \sigma(spin)=\frac{\mu ^2_r }{m^2_p} \frac{\sigma_N(spin)}{3}
  \left [\Omega_p -\Omega_n \right ]^2 F_{11}(q),\sigma(coh)=\frac{\mu ^2_r }{m^2_p}
   \sigma_n(coh) N^2  \left [ F(q) \right ]^2
  \end{equation}
  where $\Omega_p$ and $\Omega_n$ are the nuclear spin ME associated with the proton and neutron component
  and $F_{11}(q)$ is the spin response function, $N$ is the neutron number and $F(q)$
  the nuclear form factor \cite{JDV06}.
\item The Intermediate Higgs scalar

As in the neutralino case we find that, unlike the naive
expectations \cite{AGSER05} quarks other than $u$ and $d$
dominate. One finds:
   \begin{equation}
  \sigma_N(coh)=\frac{8}{\pi}\left (G_F m^2_p \right )
   ^2 \frac{m_p^2 (m_{\nu^{(1)}})^2}{m^4_h}m_p^{-2}(\sum_q f_q)^2
  \end{equation}
 Using the value we obtain $\sum_q f_q=0.67$ we obtain the results shown in
Fig. \ref{fig:nuh}. We see that this mechanism excludes a heavy
neutrino as a WIMP candidate, unless the Higgs mass is very large.
In the Standard Model this is possible and $m_h$ can be treated as
a parameter to be extracted from the data. In SUSY models,
however, the lightest neutrino is expected to be quite light,
$m_h< 120$ GeV.
\begin{figure}[!ht]
\begin{center}
\includegraphics[scale=0.6]{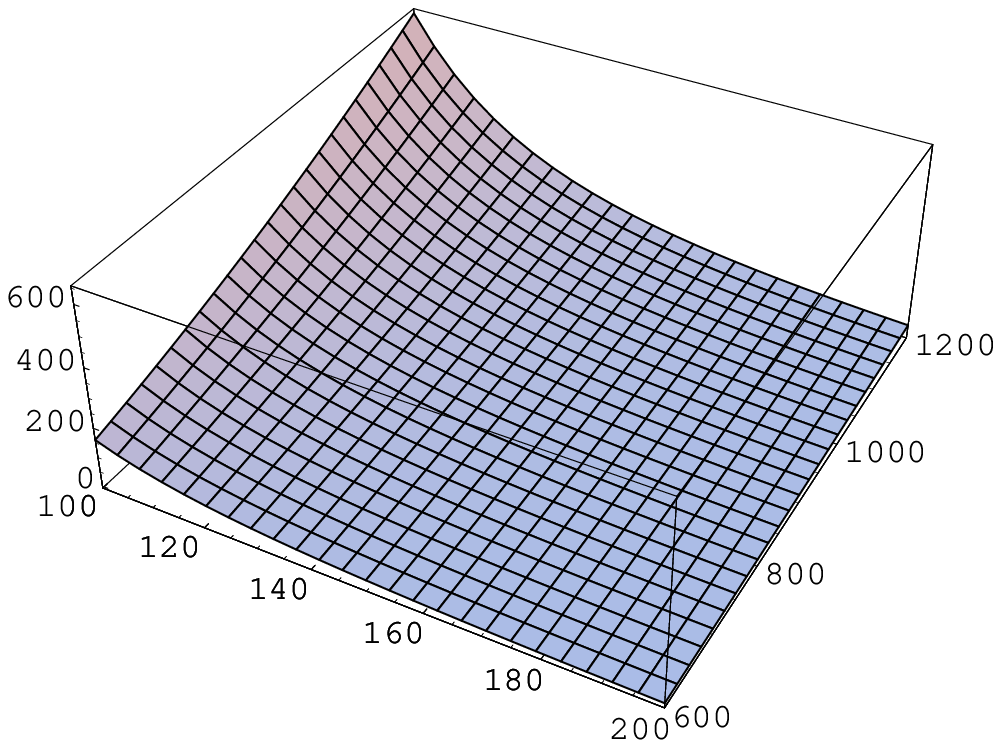}
\ \ \ \ \ \
\includegraphics[scale=0.6]{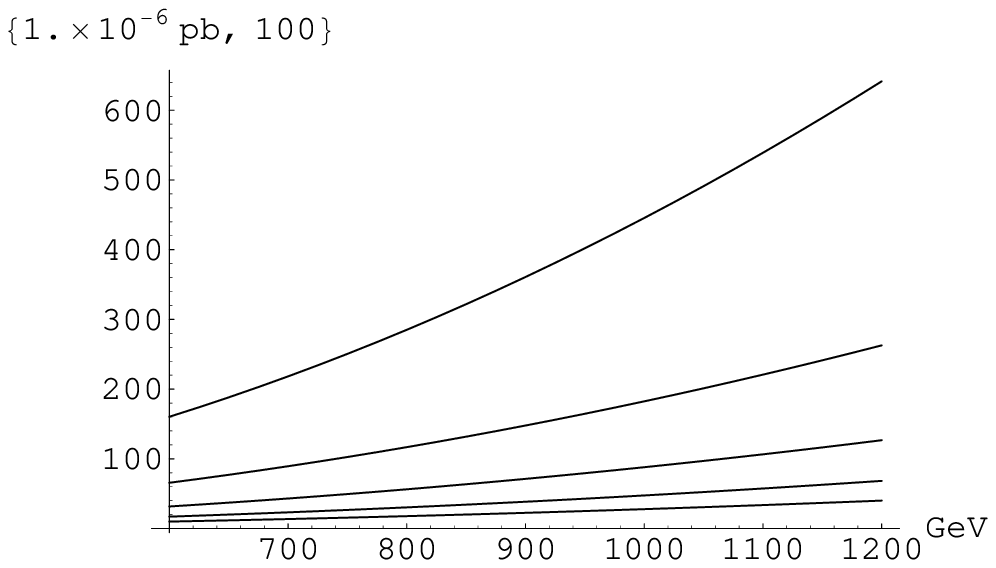}
 \caption{On the left we show
the coherent nucleon cross section as a function of
$m_{\nu^{(1)}}$ and $m_h$. On the right we show the same thing as
a function of the mass of $\nu^{(1)}$ for the indicated Higgs mass
(from top to bottom $100,125,150,175$ and $200$ GeV).
}
 \label{fig:nuh}
\end{center}
  \end{figure}
\end{enumerate}
\section{Other non SUSY Models}
\label{Zmodel}
 There exist models less well motivated than those based on K-K theories, but with a more
  restricted
particle content. We will mention two possibilities:
\begin{enumerate}
\item Extensions of the standard model \cite{MA06}.\\
Instead of the K-K parity one introduces a discrete symmetry
giving rise to a new "parity".  The lightest of added exotic
particles, the right handed neutrino and a scalar isodoublet
$\eta$, with parity opposite to the Standard Model  particles, is
cosmologically stable.
If the  lightest particle (WIMP) is the heavy neutrino, the
obtained cross sections are similar to those discussed in the
previous section in connection with $\nu^{(1)}$.
If the WIMP is the neutral component of the exotic Higgs scalar
$\eta$,
the obtained results are similar to those obtained in the previous
section involving $B^{(1)}$ and $h$, except that one encounters an
unknown quartic coupling $\lambda_{eff}$.
  \item Technicolor theories \cite{KOUVARIS06}.\\
  In this case the WIMP is the LTB (lightest neutral
  technibarion). This scalar couples to the quarks via derivative
  coupling and Z exchange. Again the only parameter is the LTP
  mass. The obtained nucleon cross sections are three times larger than  those of the K-K
  case with Dirac $\nu^{(1)}$ and lead to coherence in the number of neutrons.
\end{enumerate}
\section{Discussion}
 We will
 concentrate here on K-K particles, but our conclusions hold for all
 heavy  WIMPs. They are as follows:
\begin{itemize}
\item The K-K neutrinos as CDM candidate.\\
In this case everything is under control, except, of course, the
fact that we do not know for sure whether the K-K neutrinos are
Majorana  or Dirac particles. Most authors expect them to be Dirac
neutrinos (see, e.g., Servant\cite{SERVANT}). The Dirac neutrino
case is, however, excluded from the data, since the cross section
is too large. So we will consider the case of Majorana neutrinos.
Even in this case the Higgs contribution to the nucleon cross
section, which is proportional to the $ [m_{\nu^{(1)}} ]^2$, is
too large and excludes the K-K neutrino as a viable WIMP
candidate, unless the lightest Higgs is much heavier. In all other
cases
 the rate will
scale as in Eq. (\ref{eq:rate}). Majorana spin cross sections such
as those found above  may not be excluded from the data, since the
experimental limit on the spin cross section is much weaker than
that associated with the coherent mode and it  depends on nuclear
physics. \item The K-K boson as CDM candidate.
\begin{enumerate}
\item The only unknown parameters of the theory are the masses of
K-K quarks and gauge bosons as well as the mass of the neutral
Higgs. \item In the spin independent mechanism the proton cross
section is dominant. The event rate will be down by a factor
$\frac{Z^2}{A^2}$ compared to the analysis of the neutralino case.
 This prediction can be consistent with the present data only
away from the resonance and/or large K-K gauge boson masses.
\item Since the spin nucleon cross section is not much larger than
that of the coherent mechanism, it does not seem likely  to be
observed.
\end{enumerate}
\end{itemize}

\section*{Acknowledgments} One of the authors (JDV) is indebted to
Dr G. Servant for discussions during his visit at CERN, which was
supported by PYTHAGORAS-1, Operational Program for Education and
Initial Vocational Training of the Hellenic Ministry of Education
under the 3rd Community Support Framework and the European Social
Fund.

\end{document}